\documentstyle[12pt]{article}
\global\arraycolsep=2pt

\newcommand{\be}{\begin{eqnarray}}
\newcommand{\ee}{\end{eqnarray}}
\newcommand{\la}{\langle}
\newcommand{\ra}{\rangle}

\begin{document}

\begin{titlepage}
\begin{flushright}
SMU-HEP-94-10\\
hep-ph/9502258 \\
January 1995
\end{flushright}

\vspace{0.3cm}
\begin{center}
\Large\bf
  Chiral Condensate, Master Field
       and all that in $QCD_2 (N\rightarrow\infty)$.\\
   \end{center}

\vspace {0.3cm}

 \begin{center} {\bf Boris Chibisov\footnote{e-mail
address:chibisov@mail.physics.smu.edu}
  and Ariel R. Zhitnitsky\footnote{
On leave of absence from Budker Institute of Nuclear Physics,\\
Novosibirsk,630090,Russia.
e-mail address:arz@mail.physics.smu.edu,  }}
 \end{center}

\begin{center}
{\it Physics Department, SMU , Dallas, Texas, 75275-0175}

\end{center}
\begin{abstract}

 We discuss
the various aspects of two-dimensional $QCD_{2}(N\rightarrow\infty)$
(the 't Hooft model\cite{Hooft}).
Our main interest (motivated by the corresponding analysis in the
four dimensional QCD) is    the vacuum structure
of the theory. We use the very
 general methods in the analysis,
such as    dispersion relations and
duality
in order to relate the known spectrum of $QCD_2$ to the
different vacuum characteristics.

 We explicitly calculate
(in terms of physical parameters like masses   and matrix elements)
the chiral condensate as well as
the mixed   vacuum condensates:   $$\la 0|\bar{q}(g\epsilon_{\mu\nu}
G_{\mu\nu})^nq |0\ra \sim M_{eff}^{2n}\la 0|\bar{q} q |0\ra .$$

We interpret the factorization property for the mixed vacuum condensates
 as a  reminiscent of the master field at large $N$.

\end{abstract}
\end{titlepage}
\vskip 0.3cm
\noindent
{\bf 1. Introduction}
\vskip 0.3cm
In this paper we analyze
the vacuum properties of two-dimensional QCD. The standard way of doing
so is the ``canonical" (or non-canonical) gauge fixing, analysis of
different constraints, elimination of redundant degrees of freedom
e t.c.
We take the opposite approach. We assume that the spectrum is known
and our main goal is find out the vacuum properties which would
correspond to the  given spectrum.

Since the pioneering paper by 't Hooft\cite{Hooft}
in 1974, $QCD_2(N\rightarrow\infty)$ has been subject of many investigations
 \cite{Callan} - \cite{Kent}. The list of problems discussed in these
papers is  impressive: scaling and $e^+e^-$ annihilation;
deep-inelastic scattering and Drell-Yan formula; Regge behavior,
 analysis of form factor and wide angle behavior of exclusive amplitudes;
light cone quantization and e t.c..

The next step which has been taken
is  the analysis of the vacuum properties which would
correspond
to the  well established spectrum. It has been realized that
the vacuum of the theory in the 't Hooft
limit
\be
\label{1}
 g^2N\sim const. ~~~N\rightarrow\infty,~~~m_q
\gg g\sim\frac{1}{\sqrt{N}}
\ee
is quite nontrivial. In particular, the chiral condensate
$\la \bar{q}q\ra=-N\sqrt{\frac{g^2N}{12\pi}}$ in the
limit $m_q\rightarrow 0$ has been  calculated \cite{ARZ}\footnote{The
chiral limit $m_q\rightarrow0$ can be considered
only after the limit $N\rightarrow\infty$ is taken.These limits
do not commute, see the next section for explanation.}.The result was
confirmed by numerical \cite{Ming1},\cite{Ming2} and
independent analytical
calculations\cite{Lenz}. Moreover, the method has been  generalized for
the nonzero quark mass and the corresponding explicit formula
for the chiral  condensate $\la \bar{q}q\ra$
with arbitrary $m_q$ has been obtained
\cite{Mattis}. Let us note that there is no contradiction with the Coleman
theorem \cite{Coleman} at this point, because the BKT
(Berezinski-Kosterlitz-Thouless) behavior takes place in the
large $N$ limit\cite{ARZ}.

Recently there has been a renewal of interest in the study of $QCD_2$
\cite{Gross}-\cite{Wadia}. The motivation for this interest was
quite different for different people: it ranged from an attempt to find a
string representation of QCD in four dimensions to mastering
the instantons and master field in this theory.
The important lesson to be learned from the new development, we believe, can
be formulated   as follows:
the structure of  gauge theories even in two dimensions (with
and   without
matter fields) is very complicated. However, some physical
questions can be formulated irrespective of number of
dimensions. In certain circumstances , the correctly formulated
questions can even be answered.
This gives some hope (and hints)
that the  similar formulation of the analogous problems
  hopefully can be found
in four dimensions.

This paper is largely motivated by analysis
of nonperturbative wave functions with a minimal
number of constituents in four dimensional QCD.
As is known, such a function gives    parametrically leading contributions
to hard exclusive processes.  The corresponding
wave functions within QCD  have  been introduced into the theory
 in the late seventies
and  early eighties  \cite{Brod}  to describe the
exclusive processes. We refer to the review papers
\cite{Cher},\cite{Brod1}
 on this subject for the  detail definitions
and discussions in the given context of Wilson operator
expansion.

The two-dimensional $QCD_2(N\rightarrow\infty)$ is the {\bf ideal
toy model} for such kind of problems, because
 the only physical states in the model are
states with minimal number of constituents
(the $\bar{q}q$ mesons)\footnote{
Let us remind that an additional creation of $\bar{q}q$ pairs
is suppressed by a factor $1/N$.}. The properties of these,
nonperturbative wave functions might be quite nontrivial
and unexpected. It would be very interesting to
check the analytical nonperturbative methods (mainly the QCD sum rules)
 which have been developed
for such an analysis.

As is known, the QCD sum rules approach operates with objects like
vacuum condensates of different local operators. Eventually, if we knew
  all types of vacuum condensates, we would calculate an
arbitrary correlation function and thus, through the dispersion
relation, we would find the spectrum and amplitudes.

The subject of the present paper is analysis of the vacuum structure. More
specific, we are interested in the exact calculation of the chiral,
mixed vacuum condensates
 $\la 0|\bar{q}(g\epsilon_{\mu\nu}
G_{\mu\nu})^nq | 0 \ra  $.
Once these nonperturbative condensates are found, than
some  more complicated nonlocal vacuum expectation values
(like Wilson line
$  \la 0| W(x) |0\ra \equiv
\la 0|\bar{q}(x) e^{ig\int_0^xA_{\mu}dx_{\mu}}q(0) |0\ra $
and its superpositions $  \la 0| W(x_1) W(x_2)W(x_3)... |0\ra$)
can be in principle evaluated.

The paper is organized as follows. In Sect.2 we review
some  $QCD_2(N\rightarrow\infty)$ properties with emphasis
on the vacuum structure of the theory.
We review  the spectrum found by 't Hooft and explain why this spectrum
{\bf unavoidably leads} to the  existence of the chiral condensate $\la
\bar{q}q\ra$.
 Through the dispersion
relation we explicitly  express the condensate in terms of spectrum
and  physical matrix elements.
We qualitatively explain why the
nonzero magnitude for  $\la \bar{q}q\ra$
does not contradict to the Coleman theorem
and why the chiral limit $m_q\rightarrow  0$ can be considered only after
the large $N$ limit is taken. In different words, we demonstrate that
 to preserve the 't Hooft solution (which
corresponds to the selection only planar, leading at large $N$,
diagrams) we need to require that the inequality $m_q\gg g\sim 1/\sqrt{N}$
is fulfilled. Otherwise, the nonplanar diagrams which
might have the singular factors like $1/m_q$ come into the game and destroy
the 't Hooft solution.

Sect.3 is devoted to the calculation of the mixed vacuum
condensates. We try to avoid, in all cases, the
discussion of such complicated problems as infrared
and ultraviolet regularization, the problem of splitting
of two local operators, and many other problems which unavoidably
accompany any calculation in local field theory. Instead,
we use the knowledge of  existence
of the nonperturbative vacuum condensate $\la \bar{q}q\ra$
to calculate the more complicated, higher dimensional
condensates\footnote{ The gluon in $QCD_2$ is not
a physical degree of freedom. Thus, one could anticipate that
any local operator can be expressed in terms of the quark fields.
As   is expected, this is indeed the case.}.
The  {\bf crucial reason} for such incredible
simplification is the factorization property
 $\la Q_1\cdot Q_2\ra=\la Q_1\ra \la Q_2\ra +0(1/N)$
of vacuum expectation values in the large $N$ limit.
Thus, all vacuum condensates can be reduced to
$\la \bar{q}q\ra$ and its powers.
 We believe that if the master field
is found, it should reproduce the vacuum properties mentioned
above.

Sect.4 is our conclusion and outlook.
To be more specific, we discuss   some phenomenological
consequences of the obtained formulae
which might be interesting for the analysis of nonperturbative
 wave functions
and heavy-light quark system.
\vskip .3cm
   \noindent
{  \bf 2. The spectrum and chiral condensate in $QCD_2(N\rightarrow\infty)$.}
  \vskip .3cm

The model we shall consider consists of quark in fundamental
representation interacting via an $SU(N)$ color gauge group. We follow
the notation of ref.\cite{Callan} and present the 't Hooft equation
\cite{Hooft} in the following form:
\be
\label{2}
m_n^2\phi_n(x) =\frac{m_q^2}{x(1-x)}\phi_n(x)
-m_0^2P\int dy\frac{\phi_n(y)}{(x-y)^2},
\ee
where symbol $P$ notes as the principal value of the integral, and
$0<x<1$ is the  fraction of the total  momentum
of the bound state  carried  by quark $q$
with mass $m_q$.
The quantity
$m_0^2\equiv \frac{g^2N}{\pi}$ is the basic mass scale in the theory
and the index  $n$ classifies the ordering  number of the bound states
$|n,p\ra$
with total momentum $p_{\mu}$.
The same wave function
can be expressed in terms of the following matrix element \cite{Brower1}:
\be
\label{3}
\phi_n(x)=\sqrt{\frac{N}{\pi}}\int d y_+e^{-iy_+(1-2x)p_-}
\la 0|\bar{q}(-y)q(y)|n,p\ra|_{y_-=0}.
\ee
Let us note that matrix element on the right hand side
is written in the light cone gauge $A_-=0$; to restore
the  manifest  gauge invariance one can insert
the standard exponential factor $e^{ig\int A_-dy_+}$
into the  formula (\ref{3}).

Let us review some important properties of  equation (\ref{2}).
The entire spectrum is discrete
and classified by the integer number $n$. The wave functions $\phi_n(x)$
are orthogonal, complete and obey the following boundary conditions
\be
\label{4}
\phi_n(x)\rightarrow [x(1-x)]^{\beta},~~~x\rightarrow 0,~~x\rightarrow 1,~~~
\pi\beta\cot(\pi\beta)=1-\frac{m_q^2}{m_0^2}.
\ee

It turns out the spectrum of states is almost linear for large $n$:
\be
\label{5}
m_n^2\simeq \pi^2m_0^2 n,~~\phi_n(x)\simeq\sqrt{2}\sin(\pi n x)
\ee
and does not depend on mass of the quark.
What is more important, in the chiral limit $m_q\rightarrow 0$
the lowest level (we call it $\pi$ meson) tends to zero
$m_{\pi}^2\sim m_q$ and one could expect the nonzero
magnitude for the chiral  condensate.

We define the chiral condensate in the current algebra terms
as follows:
\be
\label{6}
0= \lim_{p_{\mu}\rightarrow 0} i\int d^2x e^{ipx}\partial_{\mu}
\la 0|T\{\bar{q}\gamma_{\mu}\gamma_{5}q(x),~
\bar{q}\gamma_{5}q(0)\}|0\ra =   \\   \nonumber
 2i\la 0|\bar{q}q|0\ra + 2m_q \cdot
\la 0|T\{\bar{q}i\gamma_{5}q(x),~
\bar{q}i\gamma_{5}q(0)\}|0\ra .
\ee
As we already mentioned, the only states  of
 't Hooft's solution are the    quark-antiquark bound  states.
Thus, they must saturate the dispersion relation. Upon inserting
this complete set of mesons to the (\ref{6}) one thus obtains:
\be
\label{7}
\la 0|\bar{q}q|0\ra =-m_q\sum_{n}\frac{N}{\pi}\frac{f_n^2}{m_n^2},
\ee
where $f_n$ is defined in terms of the following matrix elements
\be
\label{8}
\la 0|\bar{q}i\gamma_{5}q  |n\ra
=\sqrt{\frac{N}{\pi}}f_n
 ,~~f_n=\frac{m_q}{2}\cdot\int_0^1\frac{\phi_n(x)}{x(1-x)}dx
\ee
In the chiral limit the only   state which
 can contribute
to the formula (\ref{7}) is the $\pi$ meson.
 Its matrix element can be calculated exactly
and we end up with the following expression for the chiral condensate
in the $m_q\rightarrow 0$ limit \cite{ARZ}:
\be
\label{9}
\la 0|\bar{q}q|0\ra =-N\frac{m_0}{\sqrt{12}},~~m_0^2=\frac{g^2N}{\pi},
{}~f_{n=0}=\frac{m_0\pi}{\sqrt{3}},~~m_{\pi}^2=m_q\cdot\frac{2m_0}{\sqrt{3}}.
\ee
As was expected, we find that $\la 0|\bar{q}q|0\ra \sim N$.
Besides that, as we already noticed in \cite{ARZ}, if we put
$m_q=0$ from the very beginning, then $\la 0|\bar{q}q|0\ra =0$.
This  corresponds to the different regime when $m_q\ll g\sim1/\sqrt{N}$,
when nonplanar diagrams come into the game. We discuss this point
a little bit later.
The last remark is the observation that the entire nonzero answer
for the condensate comes from the infrared region of the
integration in eq.(\ref{8}): $x\sim 0, x\sim1$ which corresponds to the
situation when one of the quarks carries all the momentum and the
second one is at rest.

The sum (\ref{7}) can be calculated exactly
for arbitrary $m_q$ \cite{Mattis}. The crucial point is that
for arbitrary $m_q$
the nonzero contribution comes from the highly excited states ($
n\gg 1$) only.
The properties of these states are well-known:
\be
\label{10}
f_n^2\rightarrow \pi^2m_0^2, ~~~~~~m_n^2\rightarrow\pi^2m_0^2\cdot n,
{}~~~n\gg 1,
\ee
and thus the sum (\ref{7}) can be explicitly evaluated with the result
\cite{Mattis}:
\be
\label{11}
\la 0|\bar{q}q|0\ra=\frac{m_qN}{2\pi}
\{\log(\pi\alpha)-1-\gamma_E+(1-\frac{1}{\alpha})[I(\alpha)-
\alpha I(\alpha)-\log 4]\},
\ee
where $\alpha=\frac{m_0^2}{m_q^2},~~\gamma_E=0.5772..$
is Euler's constant and
$$ I(\alpha)=\int_0^{\infty}\frac{dy}{y^2}\frac{1-\frac{y}{\sinh y \cosh y}}
{[\alpha(y \coth y -1) +1]}.$$
This result is exact for large $N$  and arbitrary quark mass
as far as we remain in the 't Hooft regime (\ref{1}), i.e.
$m_q\gg g\sim1/\sqrt{N}.$ It reduces to the eq.(\ref{9}) in the
limit $\alpha\rightarrow\infty$, as it should.

The last condition ($m_q\gg g $ ) which has to be
 satisfied for the 't Hooft solution to be valid,
requires some additional explanation.
Roughly speaking, nonplanar diagrams may contain
a factor $\sim m_q^{-1}$ which at $m_q=0$ blow up and the theory changes
completely. The concept of the proof that there exists a factor
 $\sim m_q^{-1}$ in nonplanar diagrams is the following.

Let us consider the correlation function for $p\rightarrow 0$
\be
\label{12}
  i\int d^2x e^{ipx}
\la 0|T\{\bar{q} q(x),~
\bar{q} q(0)\}|0\ra =P(p^2 )
\ee
 The 't Hooft solution suggests that only
planar graphs   are taken into account and,
consequently, the spectral
density contains only the contribution of one meson
states. For these contributions $P_{planar}\sim N$.
In the chiral limit, we can   calculate
the  two-pion contribution  exactly!
This contribution
 is {\bf not accounted for }
in deriving (\ref{2}). Of course, the two-pion
contribution is suppressed by a factor $1/N$. However,
it contains a term $\sim\frac{m_0^2}{m_{\pi}^2}$
which tends to infinity for $m_q\rightarrow 0$.
The presence of the factor $\sim m_q^{-1}$
in nonplanar diagrams leads to the aforementioned constraint
on $m_q$ (\ref{1}).

Now, let us explicitly demonstrate the existence
of the term $\sim m_q^{-1}$ for the two-pion contribution.
In order to do so, let us write down a
 dispersion relation for $P$:
\be
\label{13}
P(0)=\frac{1}{\pi}\int_{4m_{\pi}^2}^{\infty}\frac{ds}{s}
 ImP(s) ,
\ee
where $ImP(s)$ is the physical spectral density.
The $\pi\pi$ contribution is fixed uniquely by (\ref{9}),
 because of the special
role of pions \cite{ARZ}:
\be
\label{14}
\la \pi\pi|\bar{q}q|0\ra|_{p\rightarrow0}=\frac{m_0\pi}{\sqrt{3}},~~
\frac{1}{\pi}
ImP^{\pi\pi}(s)=\frac{m_0^2\pi^2}{6}\frac{1}{\sqrt{s(s-4m_{\pi}^2)}},
\\     \nonumber
P^{\pi\pi}(0)=\frac{m_0^2\pi^2}{6}\int_{4m_{\pi}^2}^{\infty}
\frac{ds}{s\sqrt{s(s-4m_{\pi}^2)}}=
\frac{m_0^2\pi^2}{12 m_{\pi}^2}\sim\frac{1}{m_q}.
\ee
It is clear that the only cause for
a singular $\sim 1/m_q$ behavior is the
finiteness of the pion matrix elements at zero momentum.
At the same time this contribution does not contain
the  large factor $N$ which accompany a one meson contribution to the same
correlator.
To suppress these nonplanar  diagrams we need to
require $N\gg\frac{m_0^2}{m_{\pi}^2}$.
Thus, we would expect that some kind of phase transition may occur
in the region $m_q\sim g$, where we would expect a complete
restructuring of the theory.

The last subject we would like to discuss in this section
is the strict Coleman theorem  \cite{Coleman}
which states that a continuous symmetry cannot be broken
spontaneously in two dimensional theories.
As we discussed earlier \cite{ARZ} we expect that as in the
$SU(N\rightarrow\infty)$ Thirring model
(where the chiral symmetry
is ``almost" spontaneously broken \cite{Witten}),
the BKT   effect \cite{BKT} operates in regime (\ref{1}).
This fact also confirms the 't Hooft spectrum:
states with opposite $P$ parity are not degenerate
in mass and there is an ``almost" Goldstone boson
with $m_{\pi}^2\sim m_q+1/N$.

To be more specific, one can show \cite{ARZ} that in
$QCD_2 (N\rightarrow\infty)$
the behavior of the proper
two-point correlation function
is as follows:
\be
\label{15}
   \la 0|T\{\bar{q}_{L} q_{R}(x),~
\bar{q}_{R} q_{L}(0)\}|0\ra \sim x^{-\frac{1}{N}}.
\ee
Such a behavior together with claster
property at $x\rightarrow\infty$ implies
the existence of the condensate at $N=\infty$
in a full agreement with our previous discussion.
At the same time, for any finite but large  $N$, the correlator
falls off very slowly demonstrating the BKT-behavior
with no signs of contradiction to the Coleman theorem.
\vskip .3cm
   \noindent
{  \bf 3. The vacuum expectation value of the mixed
vacuum condensates.}
  \vskip .3cm

We start from the consideration of the simplest mixed vacuum
expectation value
\be
\label{17}
 \la\bar{q} D_{ \mu }D_{\nu}   q\ra=\frac{1}{2}g_{\mu\nu}
\la\bar{q} D_{ \lambda }D_{\sigma}(\gamma_{\lambda}\gamma_{\sigma}-
\epsilon_{\lambda\sigma}\gamma_5)q\ra=
\frac{1}{4}g_{\mu\nu}\la  \bar{q} ig\epsilon_{\lambda\sigma}
G_{\lambda\sigma}  \gamma_5q \ra,
 \ee
where $[D_{\mu},D_{\nu}]=-igG_{\mu\nu}$ is the  field  strength tensor of the
gauge field
and we have used the well-known properties of the $\gamma_{\mu}$ matrices
in two dimensions.
 Thus, in the chiral limit, $m_q\rightarrow 0$, the
 operator we are interested in can be expressed
exclusively in terms of the  field strength tensor
$G_{\mu\nu}$.

What's more,
 in two dimensions an arbitrary operator can be reduced
to the form which  contains
  the field
strength tensor $ig\epsilon_{\mu\nu}
G_{\mu\nu} $ only. Indeed,
the covariant derivatives
$\la\bar{q} D^n\cdot D_{ \mu }D_{\mu}   q\ra$
 placed at the very right and at the very left
(near the quark fields)
can be transformed into the operator  $ig\epsilon_{\mu\nu}
G_{\mu\nu} $ as before. To
do the same thing with operators $D_{\mu}$
 which placed somewhere in the middle,
we need to act  (say) on the right until the quark field is reached.
By doing so, step by step, we create many additional terms which
are either: commutator like
$[D_{\mu},D_{\nu}]=-igG_{\mu\nu}$ which is the
field strength operator  or
commutators like $\sim[D_{\lambda},\epsilon_{\mu\nu}
G_{\mu\nu}] $. Fortunately,   in two dimensions these terms are
related to creation of the  quark- antiquark fields
and we discard them in according to large $N$ counting rule
 \footnote{
Of course this is not the case in four dimensions
where $[D_{\mu}^2, G_{\lambda\sigma}]$ is independent
operator which can not be reduced
to some quark fields.}.

Indeed,   the quark-antiquark operator
comes   from the equation of motion
$D_{\mu}G_{\mu\nu}\sim \bar{q}\gamma_{\nu}q$
when we sum up over the common index
$\mu$. This is true in any number of dimensions.
 In two
dimensional case, even without a common index $\mu$,
the action of  $D_{\mu}$ on field strength tensor
produces the quark-antiquark operator.
 This  can be seen by noting that
in 2 dimensions
$G_{\mu\nu}(x)\sim\epsilon_{\mu\nu}E(x)$
with    scalar function $E\sim
\epsilon_{\lambda\sigma}G_{\lambda\sigma}$
and
$D_{\mu}E\sim \bar{q}\gamma_{\mu}\gamma_5q .$

 Thus, we end up with the   vacuum condensates
$\la  \bar{q}(gE)^nq  \ra $ which are
expressed exclusively in terms of the field strength tensor
and our nearest problem is their calculation.

 Here we sketch the idea of this calculation.
We   choose  the light cone gauge,
$A_{-}=\frac{1}{\sqrt{2}}(A_0-A_1)=0$.
 In this gauge we have the usual constraint in the gauge
sector (Gauss law)
(for more details  see e.g. \cite{Lenz}):
$$ \partial_{-}E^{ab}\sim g( q_+^{\dagger b} q_+^a-\frac{1}{N}
\delta^{ab}q_+^{\dagger c} q_+^c) ,$$
where $a,b $ are the color indices. Here     the
right moving fermion $q_+$ are dynamical degrees of freedom;
   the left-moving fermion  $q_-$
 are non-dynamical degrees of freedom in this gauge.
The latter can be eliminated by the following  constraint:
$$\partial_- q_- ~ \sim m_q~q_+ .$$

The next step is to use the Gauss law to calculate
the mixed vacuum condensate. For the simplest case
it looks as follows:
$$\la  \bar{q} ig\epsilon_{\mu\nu}
G_{\mu\nu} \gamma_5 q  \ra \sim
\la  q_-^{\dagger a} g E^{ab} q_+^b  \ra\sim
\la  q_-^{\dagger a} g^2 \frac{1}{\partial_-}
( q_+^a q_+^{\dagger b})q_+^b  \ra $$
Although  the  explicit expression of the
  Green function $\frac{1}{\partial_-}$
is well known
(it is the step function $\epsilon (x_{-} -x_{-}^{'})$
in the coordinate space,
see e.g.\cite{Lenz}), as will be clear soon,
we do not need its manifest form.

What does matter, is the important property of the large
$N$ limit that reduces   the expectation
value of a product of any invariant operators
to their factorized values\cite{Witten1}.
Thus, the condensate under consideration
can be presented in the following form
$$\la  \bar{q} ig\epsilon_{\mu\nu}
G_{\mu\nu}\gamma_5 q (0) \ra \sim
 g^2\la  q_-^{\dagger a}(0)q_+^a(x_{-}^{'})\ra\cdot
\la \frac{1}{\partial_-}q_+^{\dagger b}q_+^b(0)  \ra $$
 The first term on the right hand side of this expression is
  reduced to the chiral condensate
$\la \bar{q}q \ra$. Indeed,
the additional terms like $\frac{x_-^n}{n!}
\la \bar{q}(\partial^- )^n q\ra$
that come from the Taylor expansion,
are zero because of the Lorentz invariance of the vacuum state.
The second term is also reduced to the chiral condensate
if one takes into account the constraint mentioned above
and which we would  like to write down in the following
way: $\frac{1}{\partial_-} q_+ ~ \sim \frac{1}{m_q}~q_- $.

As we have already mentioned, neither explicit formula
for the Green function $\frac{1}{\partial_-}$ has been used,
nor specific regularization  has been assumed in this derivation.

The final formula for the
simplest mixed condensate (\ref{17}) which accounts for all numerical factors
takes the following form:
\be
\label{20}
\frac{1}{2!}\la\bar{q}(x_{\mu}D_{\mu})^{2}q\ra=
-\frac{1}{8}x^2\frac{ g^2\cdot\la \bar{q}q \ra^2}{m_q}.
 \ee

Few comments are in order.
First of all, as we expected, the higher dimensional condensate
can be expressed in terms of the fundamental chiral condensate
(\ref{9},\ref{11}).
We note also that the mixed condensate (\ref{20})
has the same $N$ dependence as the fundamental one,
 $\la\bar{q}q\ra$ (remember,
$g^2\la\bar{q}q\ra\sim g^2N\sim 1$).

The sign of the right hand side in eq.(\ref{20})
is the result of the calculation.   We would like to
pause here to compare this result
with somewhat analogous  in four- dimensional QCD,
where the corresponding formula looks as follows:
\be
\label{21}
\frac{1}{2!}\la\bar{q}(x_{\mu}D_{\mu})^{2}q\ra=
\frac{1}{16}x^2 \cdot\la \bar{q}ig\sigma_{\mu\nu}G_{\mu\nu}^a
\frac{\lambda^a}{2}q \ra\simeq\frac{1}{16}x^2 0.8 GeV^2\la\bar{q}q\ra.
 \ee
In both cases, the sign of the ratio
$$\frac{\la\bar{q}(x_{\mu}D_{\mu})^{2}q\ra}{\la\bar{q}q\ra}\sim ~~
x^2\frac{\la\bar{q} D_{\mu}^{2}q\ra}{\la\bar{q}q\ra}\sim x^2<0$$
is negative for negative $x^2$, where operator expansion is effective.
In four dimensions this sign has
very deep physical meaning, because there is one-to-one
correspondence between such kind of the vacuum condensates
and mean values of transverse moments for the
pion wave function\cite{Zhit1}\footnote{This is not
a big surprise, however, because the pion is Goldstone
particle and its matrix elements very often can be reduced
to the vacuum condensates through the PCAC.}. There is no
such interpretation in 2d (no transverse direction in this case),
however, one can argue that for the space-like interval,
  the sign in the eq.(\ref{20}) does correspond
to the positivity of the Hermitian operator
 $ (iD_1)^2$.

The next step is the application of the same
procedure  for the operators with
arbitrary number of gluon field insertion $E^{ab}$.
It can be done in the same way as before just
because  of the factorizability
mentioned above. Thus we
arrive to the following
formula
\be
\label{22}
\la\bar{q} (D_{ \mu }D_{\mu})^n   q\ra
=\frac{1}{2^n} \la  \bar{q}(ig\epsilon_{\lambda\sigma}
G_{\lambda\sigma}  \gamma_5)^nq \ra
 =(-\frac{g^2\la\bar{q}q\ra}{2m_q})^n\la\bar{q}q\ra ,
 \ee
which is our main result.
The   comment to this formula can be formulated
in the following way.
Each time the insertion of an  additional factor
$(D_{ \mu }D_{\mu})$, which  is proportional
to the field strength tensor $gE$, gives one and the same
numerical factor (\ref{22}). Situation can be interpreted as
we would have a {\bf classical} master field \cite{Witten1}
which we insert in place of $gE$ in the vacuum condensates.
Because of its classicality, it gives one and the same numerical factor.

The second important msg. is as follows.
The vacuum condensate of an arbitrary local operator
can be reduced through the equation of motion and
constraints to the fundamental quark condensate (\ref{9},\ref{11}).
 This is
exactly what one could expect in   $QCD_2$ where a gluon
is not a physical degree of freedom, but rather is constrained
auxiliary field which can and should be expressed
in terms of quark fields.

With these remarks in mind one could calculate
(in principle)  the vacuum expectation value
of an arbitarry nonlocal operator like string operator $\la W(x)\ra$
and its superpositions $\la W(x_1),W(x_2),W(x_3)...\ra$.
The only what we need to do is to
use the Taylor expansion
\be
\label{23}
\la W\ra=\la 0|\bar{q}(x) Pe^{ig\int_0^xA_{\mu}dx_{\mu}}q(0) |0\ra
=\sum_{n=0}^{n=\infty}\frac{1}{(2n)!}\la\bar{q}(x_{\mu}D_{\mu})^{2n}q\ra
 \ee
for the nonlocal operator we are
interested in
 and substitute the values for the vacuum condensates
we calculated previously (\ref{22}). However we better stop here.
 \vskip .3cm
   \noindent
{  \bf 5. Conclusion and Outlook.}
  \vskip .3cm
We have calculated the mixed vacuum condensates (\ref{22})
and have explained of how to reduce an arbitrary local
vacuum condensate to already known values (\ref{22}).
We also
  have demonstrated of how to
calculate  the vacuum expectation value of
the nonlocal operators (  like string operator and its superpositions)
by mean the Taylor expansion and reducing the problem
to the previous one (\ref{22}).
We shall not discuss  here the theoretical issue
 of related problems
(which are quite interesting, by the way).
Instead, in this conclusion we would like to mention
some ``phenomenological" applications which might be interesting
for the future investigation.

First of all, as we already noticed, $\la W\ra$
naturally appears in the analysis of the heavy-light quark system
$\bar{q}Q$ within operator product expansion.
 Indeed, if we consider along with \cite{Shuryak},\cite{Rad2}
 the correlation function
$\la T\{ \bar{q}Q(x),\bar{Q}q(0)\}\ra$, describing this system,
we end up (in the limit $M_Q\rightarrow\infty$) with the object
which completely factorized (in accordance with
HQET\cite{Wise})
from the heavy quark and which was called   NLC
(nonlocal condensate)\cite{Rad2}:
\be
\label{}
\la T\{ \bar{q}Q(x),\bar{Q}q(0)\}\ra
\sim  \la\bar{q}(x)P\exp(ig\int_0^xA_{\mu}dx_{\mu})q(0)\ra
+{\em perturb.~ part} .
\ee
All nontrivial, large distance physics of
the system is hidden there.
It has been noticed recently that this system might provide a
definition of the constituent quark in QCD.
Together with perturbative contribution one should expect the
following behavior for this correlator\cite{Rad2}:
\be
\label{*}
\la T\{ \bar{q}Q(x),\bar{Q}q(0)\}\ra
\sim   e^{-\Lambda \cdot x}.
\ee
 We do not expect that
the function $\la W\ra$ can
provide such a behavior by itself in t'Hooft model. The reason for that
is  related to the perturbative terms,
proportional to $\sim(\frac{g^2N}{\pi})^n(x^2)^n$.
These contributions go on the same foot as nonperturbative ones
due to the dimensional coupling constant $g$  in 2d,
and they interfere  with expansion (\ref{23}).

Naively, one could expect that nothing like that might happen in
four dimensional $QCD_4$, where the coupling constant is
dimensionless and no power corrections might occur
in perturbative series.
However this is not completely true, because of the so-called
renormalons, which may provide  some effective
power corrections \cite{Bigi}.

Another, but related issue is as follows. As is known,
\cite{Bigi2} the analysis of different problems
(like inclusive decays, distribution function...)
in heavy-light quark system, requires the precise information
about the corresponding behavior in the so-called
end-point domain.  Formally, it requires the knowledge
of the infinite set of matrix elements like
$\la H_Q|\bar{Q}D_{\mu_1}....D_{\mu_n}Q|H_Q\ra$.
Our remark is that such matrix elements
can be explicitly calculated in  $QCD_2$ as
we discussed before. We believe that
$QCD_2$   as a  toy model   may provide
at least a hint  on analytical properties
of this,  so-called universal distribution function $F(y)$
 This universal distribution function can be presented
as matrix element
$F(y)\sim
\int e^{ixy}\la H_Q|\bar{Q}(x) e^{ig\int_0^xA_{\mu}dx_{\mu}}Q(0) |H_Q\ra$.
 Operator expansion presents this (presumably regular function $F(y)$)
as a series of delta function and its derivatives
$F(y)\sim \sum a_n\delta^n(1-y)$. Thus, to find out
the analytical properties of such a function
in the real world is a difficult task
which requires the knowledge of distant terms of the series.
The lessons which might be provided by $QCD_2$ could be useful.

At the end, then, we come back to the beginning.
We have already noticed in the Introduction that $QCD_2$ is the nice
 model  for analysis of nonperturbative wave functions themselves
and methods which one can use  to extract the corresponding
information. We refer the reader
to the recent papers on this subject \cite {Zhit2},\cite{Rad1}
for the references and new development.
Here we would like   to note that in $QCD_2$ we know
the  nonperturbative wave functions
as well as whole set of local condensates (\ref{22}) which
play the crucial role in   this study.
Thus, one may try to check the methods
which have been developed for corresponding analysis.

It is a pleasure to thank Professors Gary McCartor and
Kent Hornbostel  for   encouragement
and   interesting discussions related to the subject of this paper.

This work is supported by the Texas National Research
Laboratory Commission under  grant \# RCFY 93-229.

\end{document}